\documentclass[prl]{revtex4}

\usepackage{graphicx}
\usepackage{amsmath,amssymb}
\usepackage{bm}
\usepackage{hyperref}
\usepackage{ulem}
\usepackage{slashed}
\usepackage[usenames]{color}

\hypersetup{
    colorlinks=true,
    linkcolor=red,
    citecolor=blue,
}

\def\be{\begin{equation}}
\def\ee{\end{equation}}
\def\ba{\begin{eqnarray}}
\def\ea{\end{eqnarray}}

\begin{document}

\title{
Quantum Excitations in Time-Dependent Backgrounds
}

\date{\today}

\author{Matthew Kolopanis and Tanmay Vachaspati}
\affiliation{ 
Physics Department, Arizona State University, Tempe, AZ 85287, USA.
}

\begin{abstract}
We give a technique for calculating the occupation number of quantum fields in 
time-dependent backgrounds by using the relation between one-dimensional 
{\it quantum} oscillators and two-dimensional {\it classical} oscillators.
We illustrate our method by giving closed analytical results for the 
time-dependent spectrum of occupation numbers during gravitational collapse in 
any number of dimensions. 
\end{abstract}


\maketitle


Frequently we are interested in calculating the spectrum of excitations of a quantum 
field produced due to a time-dependent background. The context can range from 
nuclear physics, condensed matter systems, gravitational systems, and cosmology,
and includes particle production during gravitational collapse or in an expanding
universe. Usually such problems are solved by the method of Bogoliubov
transformations \cite{Bogoliubov:1947,BirrellDavies}, and require knowledge of all the modes of the
quantum field at the initial and final times. Then the excitation spectrum, as well
as local information, {\it e.g.} energy-momentum densities, can be obtained. However, 
in certain 
cases it may be difficult to find all the modes of the quantum field, especially in a
time-dependent situation, and it is desirable to find a short cut to the spectrum of 
excitations, even if all the detailed local information is not available. In this paper, 
we propose such a method based on the functional Schrodinger equation that can 
be applied to certain systems. We illustrate the method in the special case 
of quantum excitations produced during gravitational  collapse.

Consider a scalar field $\Phi (t,{\bf x})$ that is free but for interaction with a time-dependent background {\it e.g.} a time-varying mass parameter or background metric. We
can always expand the field in a complete basis of functions,
\be
\Phi (t,{\bf x}) = \sum_k a_k(t) f_k({\bf x})
\ee
where $k$ denotes a complete set of mode numbers. The $f_k$'s need not be
a set of (instantaneous) mode functions; the only requirement is that any function of 
${\bf x}$ should be expressible as a linear sum of $f_k$'s.
On inserting this form in the action, we obtain
\be
S = \int dt \left [ \frac{1}{2} {\dot a}_k \bar {\bf M}_{kl}(t) {\dot a}_l 
                           - \frac{1}{2} a_k {\bf V}_{kl}(t) a_l  \right ]
 \label{SMV}
\ee
where $\bar {\bf M}$ and ${\bf V}$ are real, symmetric matrices that are given by
spatial integrals over products of the basis functions and their derivatives. They depend 
on time due to the
interaction of $\Phi$ and the background, and a sum over repeated
indices in Eq.~(\ref{SMV}) is implicit. In general, the elements of $\bar {\bf M}$ will have complicated 
time dependence and the problem appears intractable. However, progress
can be made if the time dependence can be pulled out,
\be
\bar{\bf M} = \frac{1}{B(t)} {\bf M}
\label{MM}
\ee
where ${\bf M}$ is independent of time. Such a simplification occurs for
gravitational collapse because $B \to 0$ and  then ${\bar {\bf M}}$ has the form in 
Eq.~(\ref{MM}) to leading order in $1/B$ \cite{Vachaspati:2006ki}. Similar
simplification will also occur in cosmological spacetimes in which all the time 
dependence is in an overall scale factor.

With Eq.~(\ref{MM}) one can write,
\be
S = \int d\eta \left [ \frac{1}{2}  a_k' {\bf M}_{kl}  a_l' 
                           - \frac{1}{2B(\eta)} a_k {\bf V}_{kl}(\eta ) a_l  \right ]
\ee
where primes denote derivatives with respect to a new time coordinate $\eta$
that is defined by
\be
\frac{d\eta}{dt} = B(t)
\label{detadt}
\ee

Now we attempt to do a ``principal axis transformation''.  As is standard procedure
\cite{Goldstein}, first we diagonalize ${\bf M}$. Then by a rescaling transformation, 
${\bf M}$ can be transformed into the
identity matrix. Since ${\bf M}$ is taken to be independent of time, the
diagonalization and rescaling transformations are time independent.. 
The transformed potential matrix will still 
be time dependent. The tractable case is when the transformed potential 
matrix can be diagonalized by a time independent transformation. The final action takes 
the form
\be
S = \sum_k \int d\eta \left [ \frac{1}{2}  {b_k'}^2  - \frac{v_k(\eta)}{2B(\eta)} b_k^2  \right ]
\label{finalS}
\ee
where $b_k$'s are the transformed mode coefficients and $v_k(\eta)$
are the eigenvalues of the transformed potential matrix. It should be noted that
the principal axis transformation can be implemented even if the matrices ${\bf M}$ and 
${\bf V}$ do not commute. The theorem that only commuting matrices can be
simultaneously diagonalized applies when we restrict the diagonalization procedure
to similarity transformations. The principle axis transformation involves a rescaling
which is not implemented by a similarity transformation.

We can now quantize each mode separately. The Schrodinger equation for a mode
is
\be
\biggl [
- \frac{1}{2} \frac{\partial^2}{\partial b^2}
 + \frac{1}{2} \omega^2 (\eta ) b^2 \biggr ] \psi (b,\eta )
   = i \frac{\partial}{\partial \eta } \psi (b,\eta )
\label{shostandard}
\ee
where we have omitted the mode label $k$ for convenience and written the
equation in a form reminiscent of the simple harmonic oscillator (SHO). The only 
time dependence is due to a time varying frequency $\omega(\eta)$.

The Schrodinger equation has the ``ground state'' solution \cite{Dantas:1990rk}
\begin{equation}
\psi(b,\eta) = e^{i\alpha (\eta)}
            \left ( \frac{1}{\pi \rho^2} \right )^{1/4}
            \exp \left [ \frac{i}{2}
                   \left ( \frac{\rho_\eta}{\rho} +
                           \frac{i}{\rho^2} \right ) b^2
                 \right ]
\label{psisolution}
\end{equation}
where $\rho_\eta$ denotes derivative of $\rho (\eta)$ with
respect to $\eta$, and $\rho$ is given by the real solution
of the ordinary differential equation
\begin{equation}
{\rho_{\eta\eta}} + \omega^2 (\eta) \rho = \frac{1}{\rho^3}.
\label{rhoeq}
\end{equation}
Initial conditions for $\rho$ are chosen so that the wave function yields the ground
state of the time-independent SHO at $\eta=0$,
\begin{equation}
\rho (0) = \frac{1}{\sqrt{\omega_0}}       \ , \ \ \
{\rho_\eta}(0) =0
\label{rhoinitial}
\end{equation}
where $\omega_0 \equiv \omega(0)$.
The phase $\alpha$ is given by
\begin{equation}
\alpha (\eta ) = - \frac{1}{2} \int_0^\eta  \frac{d\eta'}{\rho^2 (\eta')}
\end{equation}

The wave function in Eq.~(\ref{psisolution}) can now be expanded in 
the ``instantaneous'' SHO eigenfunctions with frequency $\omega (\eta)$.
The occupation number of the mode is given by the expectation value of
the quantum number of the excitation. The occupation number as a function 
of time and frequency can be written as \cite{Vachaspati:2006ki}
\be
N(t, \omega_0) = \frac{1}{4{\omega}\rho^2} 
                                      \left  [ ({\omega}\rho^2-1)^2 + (\rho\rho_\eta)^2 \right  ] 
\label{Nformula}
\ee
where the $\omega$ on the right-hand side refers to $\omega(\eta)$.
In (\ref{Nformula}) we have corrected the pre-factor stated in Ref.~\cite{Vachaspati:2006ki},
$1/4$ instead of $1/\sqrt{2}$ \footnote{The error was in going from Eq.~(B11) to 
(B12) of \cite{Vachaspati:2006ki}. We thank Dmitry Podolsky for bringing this 
factor to our attention.}.

The crucial step that remains is to solve Eq.~(\ref{rhoeq}) and it is here that
the remarkable connection between quantum SHO in 1 dimension and classical
SHO in 2 dimensions comes into play \cite{Lewis:1968,Lewis:1969,Brown:1991zz,
Song:2000,Parker:1971}. The solution for $\rho$ is given by
the solutions for two classical SHO's $\xi$ and $\chi$, each of which satisfies
the equation
\be
f_{\eta\eta} + \omega^2(\eta) f =0.
\label{csho}
\ee
Then the solution for $\rho$ is
\be
\rho = \frac{1}{\sqrt{\omega_0}} \sqrt{\xi^2 + \chi^2 }.
\label{rhosoln}
\ee
The initial conditions for $\rho$ can be satisfied provided
\be
\xi (0)=1, \ \ \xi_\eta (0)=0,\ \ \chi(0)=0,\ \ \chi_\eta(0) =\omega_0
\label{xichiinitial}
\ee
The time derivative of $\rho$, which occurs in the wave function,
can be written in terms of $\xi$ and $\chi$ and their derivatives,
\be
\rho_\eta = \frac{1}{\omega_0\rho} (\xi \xi_\eta + \chi \chi_\eta ) .
\label{rhoetasoln}
\ee

An explicit scheme for constructing solutions with the correct initial conditions
can now be described. We start with the classical SHO equation (\ref{csho}) and
let $f_1$ and $f_2$ be any linearly independent solutions. Then the solutions with
the correct initial conditions in Eq.~(\ref{xichiinitial}) are given by
\be
\xi(\omega_0,\eta) = W^{-1} ( f_{2\eta}(\omega_0,\eta_i) f_1(\omega_0,\eta) -
                                                   f_{1\eta}(\omega_0,\eta_i) f_2(\omega_0,\eta))
\label{xigen}
\ee
\be
\chi(\omega_0,\eta) = -\omega_0 W^{-1} ( f_{2}(\omega_0,\eta_i) f_1(\omega_0,\eta) -
                                                   f_{1}(\omega_0,\eta_i) f_2(\omega_0,\eta))
\label{chigen}
\ee
\be
\xi_\eta(\omega_0,\eta) = W^{-1} ( f_{2\eta}(\omega_0,\eta_i) f_{1\eta}(\omega_0,\eta) -
                                                   f_{1\eta}(\omega_0,\eta_i) f_{2\eta}(\omega_0,\eta))
\label{xietagen}
\ee
\be
\chi_\eta(\omega_0,\eta) = -\omega_0 W^{-1} (f_{2}(\omega_0,\eta_i) f_{1\eta}(\omega_0,\eta) -
                                                   f_{1}(\omega_0,\eta_i) f_{2\eta}(\omega_0,\eta))
\label{chietagen}
\ee
where $\omega_0 \equiv \omega(0)$ and $W$ is the Wronskian defined as
\be
W = f_{2\eta}(\omega_0,\eta_i) f_1(\omega_0,\eta_i) -
       f_{1\eta}(\omega_0,\eta_i) f_2(\omega_0,\eta_i)
\label{wronskian}
\ee
Now it is simply a matter of finding $\rho$ and $\rho_\eta$ from equations
(\ref{rhosoln}) and (\ref{rhoetasoln}), and then inserting into the occupation
number formula (\ref{Nformula}).

We will illustrate the above method by applying it to particle production of a
massless scalar field during gravitational collapse of a spherical domain wall 
in 3 spatial dimensions. The function $B(t)$ in Eq.~(\ref{MM}) is given by
 \cite{Vachaspati:2006ki}
\be
B(t) = 1-\frac{R_S}{R(t)}
\ee
where $R(t)$ is the radius of the wall at time $t$, and $R_S$ is the Schwarzschild 
radius. For convenience we will set $R_S=1$. Using the classical solution for 
$R(t)$ in the late time limit and Eq.~(\ref{detadt}), a model for
$\omega(\eta)$ can be taken to be  \footnote{The model corresponds to collapse
of an infinite spherical domain wall but with vanishing surface tension, so that
the mass is finite. An improved model starting
with a finite collapse leads to qualitatively the same results.}
\be
\omega(\eta)= \frac{\omega_0}{\sqrt{B(\eta)}} = \frac{\omega_0}{\sqrt{1-\eta}}
\ee
Then (\ref{csho}) can be solved in terms of Bessel functions {\it e.g.} see \cite{aands}. 
With the help of Eqs.~(\ref{xigen}-\ref{chietagen}), we get
\be
\xi = \frac{\pi u}{2} [ Y_0 (2\omega_0) J_1(u) - J_0(2\omega_0) Y_1(u)]
\label{xisol}
\ee
\be
\chi = \frac{\pi u}{2} [ Y_1 (2\omega_0) J_1(u) - J_1(2\omega_0) Y_1(u)]
\label{chisol}
\ee
\be
\xi_\eta = - \pi \omega_0^2 [ Y_0 (2\omega_0) J_0 (u) - J_0 (2\omega_0) Y_0(u)]
\label{xietasol}
\ee
\be
\chi_\eta = - \pi \omega_0^2 [ Y_1(2\omega_0) J_0(u) - J_1(2\omega_0) Y_0(u)]
\label{chietasol}
\ee
where $u \equiv 2\omega_0 \sqrt{1-\eta}$.
The particle occupation number per mode at time $t $ and frequency $\omega_0$ is found  
from Eq.~(\ref{Nformula}) and can be written as \cite{Vachaspati:2006ki} 
\be
N(t , \omega_0)= \frac{\sqrt{1-\eta}}{4 (\xi^2+\chi^2)} \left [ \left (
     \frac{\xi^2+\chi^2}{\sqrt{1-\eta}}-1 \right )^2 + 
     \frac{1}{\omega_0^2}(\xi\xi_\eta + \chi\chi_\eta)^2 \right ] .
\label{Nresult}
\ee
where $\xi$, $\chi$, $\xi_\eta$, $\chi_\eta$ are given in Eqs.~(\ref{xisol}), (\ref{chisol}), 
(\ref{xietasol}), and (\ref{chietasol}).

\begin{figure}[htbp]
\begin{center}
  \includegraphics[height=0.30\textwidth,angle=0]{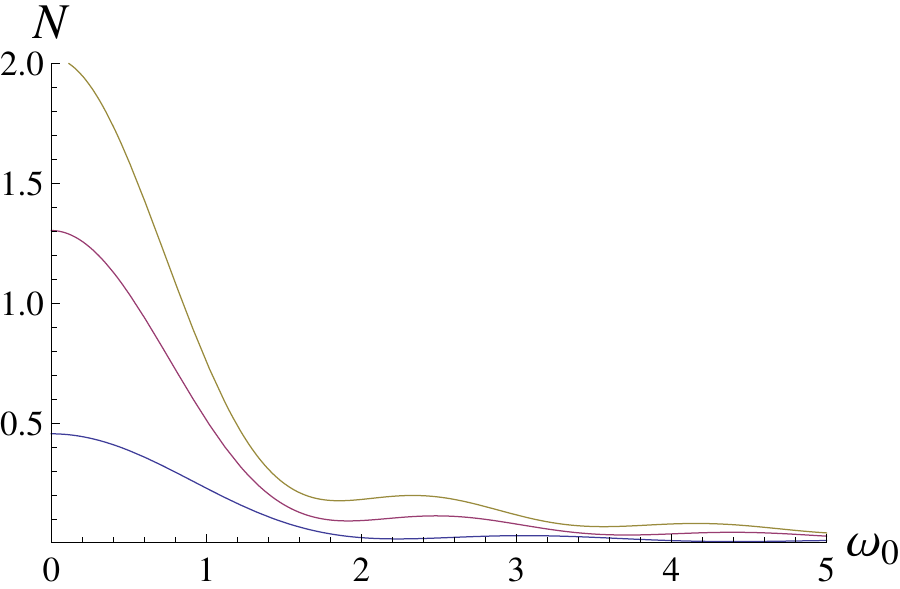}
\caption{
Occupation number, $N$, as a function of $\omega_0$ for
$\eta=0.92$, $0.98$, and $0.99$
(lowest to highest curve).}
\label{Nvsomega0}
\end{center}
\end{figure}

\begin{figure}[htbp]
\begin{center}
  \includegraphics[height=0.30\textwidth,angle=0]{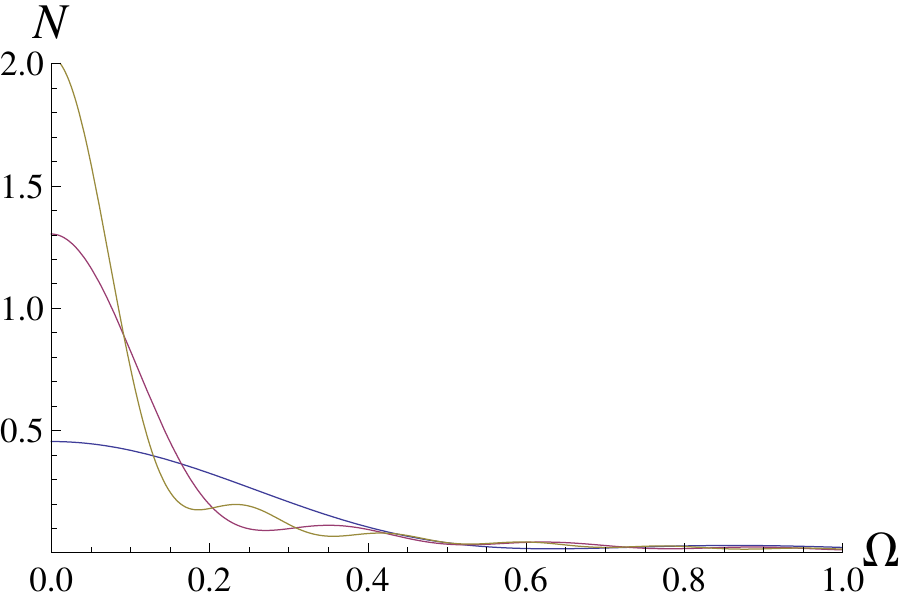}
\caption{Occupation number, $N$, as a function of $\Omega$ for
$\eta=0.92$, $0.98$, and $0.99$
(lowest to highest curve at origin).}
\label{NvsOmega}
\end{center}
\end{figure}

\begin{figure}[htbp]
\begin{center}
  \includegraphics[height=0.30\textwidth,angle=0]{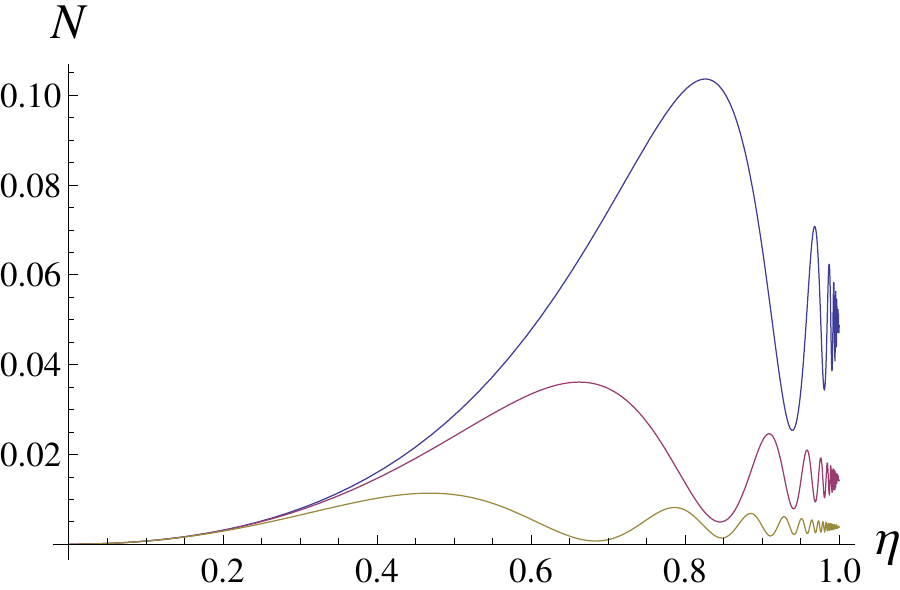}
\caption{Occupation number versus $\eta$ for $\Omega=0.5$, $1$, and $2$ (highest to 
lowest curve).}
\label{Nvseta}
\end{center}
\end{figure}

In Fig.~\ref{Nvsomega0} we plot the spectrum of occupation numbers
at three different times. The plot can also be made in terms of $\Omega$,
the final frequency corresponding to the original time parameter $t$;
$\Omega$ is also the frequency relevant for an asymptotic observer.
The relation between $t$ and $\eta$ is obtained from Eq.~(\ref{detadt}),
\be
t = \int \frac{d\eta}{B(\eta)} = \int\frac{d\eta}{1-\eta}= -\ln(1-\eta).
\ee
and the relation between $\Omega$ and $\omega_0$ is
\be
\Omega = \frac{d\eta}{dt} \omega (\eta) = e^{-t/2}\omega_0.
\ee
The occupation number spectrum in terms of $\Omega$ is shown
in Fig.~\ref{NvsOmega}, and Fig.~\ref{Nvseta} illustrates the time dependence 
of the occupation number at three different frequencies.

Limiting forms of the occupation numbers in different frequency ranges are
obtained by using expansions of the Bessel functions. The final results can
be written as
\ba
N \approx  
\begin{cases}
{e^{t/2}}/4 , \ \ \ \ & 2\Omega \ll e^{-t/2} \\
{1}/(4\pi \Omega), \ \ \ \ & e^{-t/2} \ll 2 \Omega \ll 1 \\
{1}/(64\Omega^2), \ \  \ \ & 1 \ll 2 \Omega .
\end{cases}
\label{Ncases}
\ea
It is also possible to give a closed formula for the occupation numbers for 
$t \to \infty$ that is valid at all frequencies
\be
N \to \frac{1}{4} \frac{\pi \Omega}{(J_1^2+Y_1^2)} \left [ 
 \biggl \{ J_1^2+Y_1^2- \frac{1}{\pi \Omega} \biggr \}^2 +
  \biggl \{J_0J_1+Y_0Y_1 \biggr \}^2 \right ]
  \label{Nforeta1}
\ee
where all the Bessel functions have argument $2\Omega$. 

The occupation number in the intermediate regime $e^{-t/2}=\sqrt{1-\eta}\ll 2 \Omega \ll 1$
correspond to the low frequency part of the thermal blackbody temperature
\be
N_{\rm T} = \frac{1}{e^{\Omega/T}-1}
\ee
if we take the temperature to be
\be
T = \frac{1}{4\pi}
\ee
in units where the Schwarzschild radius $R_S =1$. This temperature is precisely the 
Hawking temperature \cite{Hawking:1974sw}. The very low frequency limit in 
Eq.~(\ref{Ncases}) only coincides with the divergent thermal value if we take 
$\eta \to 1$ ($t \to \infty$). 

The high frequency limit in Eq.~(\ref{Ncases}) does not match the exponentially 
(Boltzmann) suppressed thermal expression. This can be understood by noting
that our calculation yields occupation numbers during the collapse and includes
all excitations that are produced, including transients and non-radiative excitations
(vacuum polarization). Such non-thermal features are also seen in a treatment of 
gravitational collapse in a 1+1 dimensional model \cite{Davies:1976ei,Davies:1976} 
and are separated out by using the explicit mode functions to calculate the energy flux 
at spatial infinity \footnote{Non-thermaility of the occupation numbers is also evident if we calculate
fluctuations of the occupation numbers. We find 
$\langle N^2 \rangle - \langle N\rangle^2 = 2 \langle N\rangle (\langle N\rangle+1)$,
a relation that is independent of $\omega_0$. The corresponding expression for
thermal fluctuations does not have the factor of 2 on the right-hand side. We
thank Rong Chen for his help in deriving this result.}.
Only the radiative part, {\it i.e.} the Hawking radiation, is thermal and 
hence exponentially suppressed at high frequencies. If we are interested in asymptotic
signatures of gravitational collapse, it is indeed only the radiative excitations that are
of interest. However, if we are interested in the dynamics of the collapse, then
it is necessary to consider all modes of the scalar field and the full wave function as given in Eq.~(\ref{psisolution}).

The above comments highlight the advantages and limitations of the
technique developed in this paper. For systems in any number of 
dimensions that have the simplifying properties discussed around Eqs.~(\ref{MM})
and (\ref{finalS}) we can obtain the complete wave function for the mode 
coefficients. The solution for the wave function, Eq.~(\ref{psisolution}), completely 
characterizes the system and allows us to calculate the occupation numbers analytically 
and in closed form, even without explicit knowledge of the mode functions.
However, it becomes necessary to find the mode functions if one requires
spatial information about the excitations that are produced. Additionally, we have 
neglected backreaction of the radiation on the collapse. This is a difficult problem 
that has not seen any clear resolution in the literature; some attempts in the present 
formalism that lead to a modification of the wave function describing gravitational
collapse may be found in \cite{Vachaspati:2007hr,Vachaspati:2007ur}.

\acknowledgments
We thank Rong Chen, Yi-Zen Chu, Paul Davies, and Dejan Stojkovic for 
discussions. This work was supported by the DOE at ASU. 



\begin{thebibliography}{99}

\bibitem{Bogoliubov:1947}
N.~Bogoliubov, J. Phys. (USSR) 11, 23 (1947).

\bibitem{BirrellDavies}
``Quantum Fields in Curved Space'',
N.~Birrell and P.C.W.~Davies, Cambridge University Press (1994).

\bibitem{Goldstein}
``Classical Mechanics'', H.~Goldstein, Addison-Wesley (1980).

\bibitem{Dantas:1990rk} 
  C.~M.~A.~Dantas, I.~A.~Pedrosa and B.~Baseia,
   Phys. Rev. A {\bf 45}, 1320 (1992).

\bibitem{Vachaspati:2006ki} 
  T.~Vachaspati, D.~Stojkovic and L.~M.~Krauss,
  Phys.\ Rev.\ D {\bf 76}, 024005 (2007)
  [gr-qc/0609024].

\bibitem{Lewis:1968}
H.R.~Lewis, Jr, J. Math. Phys. {\bf 9}, 1976 (1968).

\bibitem{Lewis:1969}
H.R.~Lewis, Jr. and W.B.~Riesenfeld, J. Math. Phys. {\bf 10}, 1458 (1969).

\bibitem{Brown:1991zz} 
  L.~S.~Brown,
  Phys.\ Rev.\ Lett.\  {\bf 66}, 527 (1991).
  
 \bibitem{Song:2000}
  D-Y. Song, Phys. Rev. A {\bf 62}, 014103 (2000).
  
\bibitem{Parker:1971}
  L. Parker, Am. J. Phys. {\bf 39}, 24 (1971).
  
 \bibitem{aands}
``Handbook of Mathematical Functions'',
M. Abramowitz and I. A. Stegun, National Bureau of Standards,
Applied Mathematics Series - 55, Washington D. C. (1972);
http://people.math.sfu.ca/~cbm/aands/

\bibitem{Hawking:1974sw} 
  S.~W.~Hawking,
  Commun.\ Math.\ Phys.\  {\bf 43}, 199 (1975)
  [Erratum-ibid.\  {\bf 46}, 206 (1976)].

\bibitem{Davies:1976ei} 
  P.~C.~W.~Davies, S.~A.~Fulling and W.~G.~Unruh,
  Phys.\ Rev.\ D {\bf 13}, 2720 (1976).

\bibitem{Davies:1976}
P.~C.~W.~Davies,
Proc.\ R.\ Soc.\ Lond.\ A {\bf 351}, 129 (1976).

\bibitem{Vachaspati:2007hr} 
  T.~Vachaspati and D.~Stojkovic,
  Phys.\ Lett.\ B {\bf 663}, 107 (2008)
  [gr-qc/0701096].

\bibitem{Vachaspati:2007ur} 
  T.~Vachaspati,
  Class.\ Quant.\ Grav.\  {\bf 26}, 215007 (2009)
  [arXiv:0711.0006 [gr-qc]].



\end{thebibliography}
\end{document}